# In a nanoscience lab

Pascale Bayle-Guillemaud[a], Emmanuel Hadji[a], Peter Reiss[b], Jacques Villain[c]

*a*. SP2M, UMR-E CEA / UJF-Grenoble 1, INAC, F-38054 Grenoble, France

*b*. SPrAM, UMR CEA / CNRS / UJF-Grenoble 1, INAC, Grenoble, F-38054, France

*c*. Theory group, ESRF, B.P. 220, F-38043 Grenoble Cedex 9

*Abstract*

The production, observation and manipulation of very small objects is a tour de force, but these objects, which could infiltrate anywhere without being seen, may arouse suspicion. To assess the situation at best, we describe the activity of a nanoscience research institution, some of the methods used there, the spirit of its researchers and their attitude towards risk.

Dans un laboratoire de nanosciences. *La fabrication, l'observation et la manipulation d'objets très petits est un tour de force, mais ces objets, susceptibles de s'infiltrer partout sans être perçus, peuvent susciter la méfiance. Pour mieux apprécier la situation, nous décrivons l'activité d'un institut de recherche spécialisé dans les nanosciences, certaines des méthodes qui y sont utilisées, l'esprit de ses chercheurs et leur attitude vis à vis des risques.*

## **Introduction**

Since the beginning of this century, research has evolved considerably in the field of condensed matter, and nanoscience emerged as a new field of research. It is now one of the main activities in many institutions. In this article we shall visit one of them, the campus MINATEC in Grenoble. What is being done in such a complex? How does one come to make, observe, manipulate so tiny objects? This is a bit mysterious to many, including many physicists. It therefore seemed useful to ask some experts to explain in simple terms how they are doing, what are their purposes, and what place is given in their professional life to the new risks associated with these new objects.

The researchers who appear in this article belong to three laboratories. In one of these laboratories nano-objects are fabricated by physical methods of crystal growth, such as molecular beam epitaxy. In the second one, different nano-objects are produced, in larger amounts, by chemical methods, combining reagents in solvents. In the third laboratory, the obtained structures are observed, and sometimes they are manipulated. These examples are not intended to give a comprehensive description of research in nanoscience, but rather to give a vivid and concrete picture of research.

### **Manufacturing nanoobjets by physical methods: the case of nanowires.**
Nanowires are one example of the nano-objects made in these laboratories. They are single crystals whose diameter may be a few tens of nanometers while the length can reach several micrometers. Figure 1, obtained by scanning electron microscopy, shows how perfect the faces may be, and how uniform the diameter. The laws of surface tension are responsible for this perfection. But these same laws, which tend to minimize the surface of solids, should oppose the growth of wire-like objects. How can such objects then be obtained? For the sake of simplification, we shall assume that the wire is made by depositing a material (adsorbate) on the surface of another material (substrate). In some cases, the deposit is possible only through a catalyst, typically gold. The first step is then to deposit nanometer sized droplets of gold on the substrate and then introduce the adsorbate in gaseous form. It sticks to the substrate under the gold droplets, thus formimg columns that grow, wearing their golden head (Fig. 1a). The (nanometric) size of the droplets determines the diameter of the columns. The theory of the phenomenon is described by Dubrovskii et al. [1].

However, nanowires can sometimes grow without catalyst. Examples are zinc oxide (ZnO) nanowires on sapphire (Fig. 1c, d). The phenomenon, which occurs in quite specific conditions (well-defined substrate temperature and molecular beam intensity) was analyzed by Périllat-Merceroz *et al*. [2] but is not fully understood. Note that the ZnO crystal, unlike many conventional semiconductors, is not cubic but hexagonal. The highly anisotropic observed shape is therefore not so surprising if the surface energy and/or the growth rate are lower parallel to the hexagonal axis than in the perpendicular directions.

The realization of these objects is already an achievement, but constitutes only the first step in a program. The next step is to analyze them, i.e. to measure their physical properties (electrical conductivity, optical properties etc..) which are different from those of a bulk sample. This requires observing and manipulating nano-objects.

Manipulation is also necessary to use the nano-objects, e.g. for integration in nano-electronic devices where their purpose is to transport, either electrons or photons. The reduction in volume allows to make more sophisticated instruments in a given volume, and also to reduce heating. But in the research laboratory that we visit, the fabrication and study of the physical properties pass before applications. The knowledge obtained from this study is a prerequisite for the development of technological applications, which, however, may require to adapt the methods of fabrication.

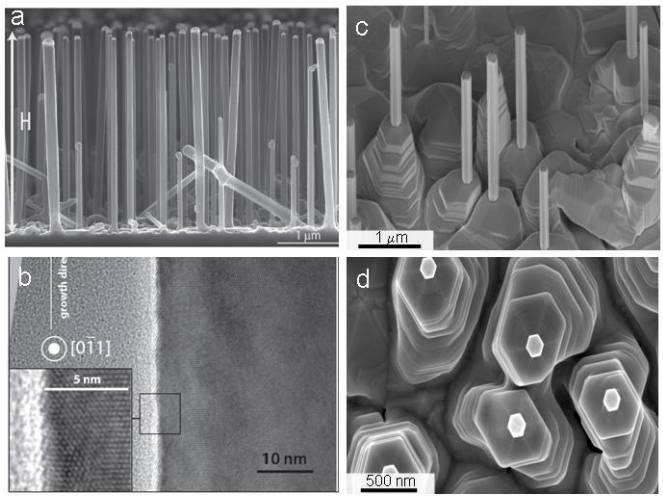

Figure 1: a) Image obtained by scanning electron microscopy (SEM) of a forest of Si nanowires with catalysts at the top; b) atomic resolution image of the edge of a nanowires obtained by transmission electron microscopy (TEM) ; c) and d) SEM image of ZnO nanowires without a catalyst obtained on sapphire. We are grateful to Pascal Gentile (INAC/SP2M/SiNaPS) for providing us with Fig. 1a and to P.H. Jouneau for Fig. 1d. Fig. 1c is taken from ref. 2 with the kind permission of the publisher and the authors.

Observation and manipulation of nano-objects.

The observation of nanometer-sized objects requires suitable instruments that can see, count and locate these objects. The scanning electron microscope (SEM) as that of Figure 2 (left), is the ideal instrument. Through an electron beam which scans the surface, it produces an image with a resolution of one nanometer. A more detailed information (on the scale of the atom, i.e. 0.1nm) can be obtained from a transmission electron microscope like the one in Figure 2 (right). The electron microscope, with its resolving power, will thus replace the eye of the researcher in nanoscience. It allows to manipulate a nanowire, to drop it, to detach it from its surface, or even move it onto another substrate for a measurement. In the absence of clamps, rudimentary nanomanipulators can be used, for instance nanometric spikes driven by piezoelectric components which can move them in a strictly controlled way.

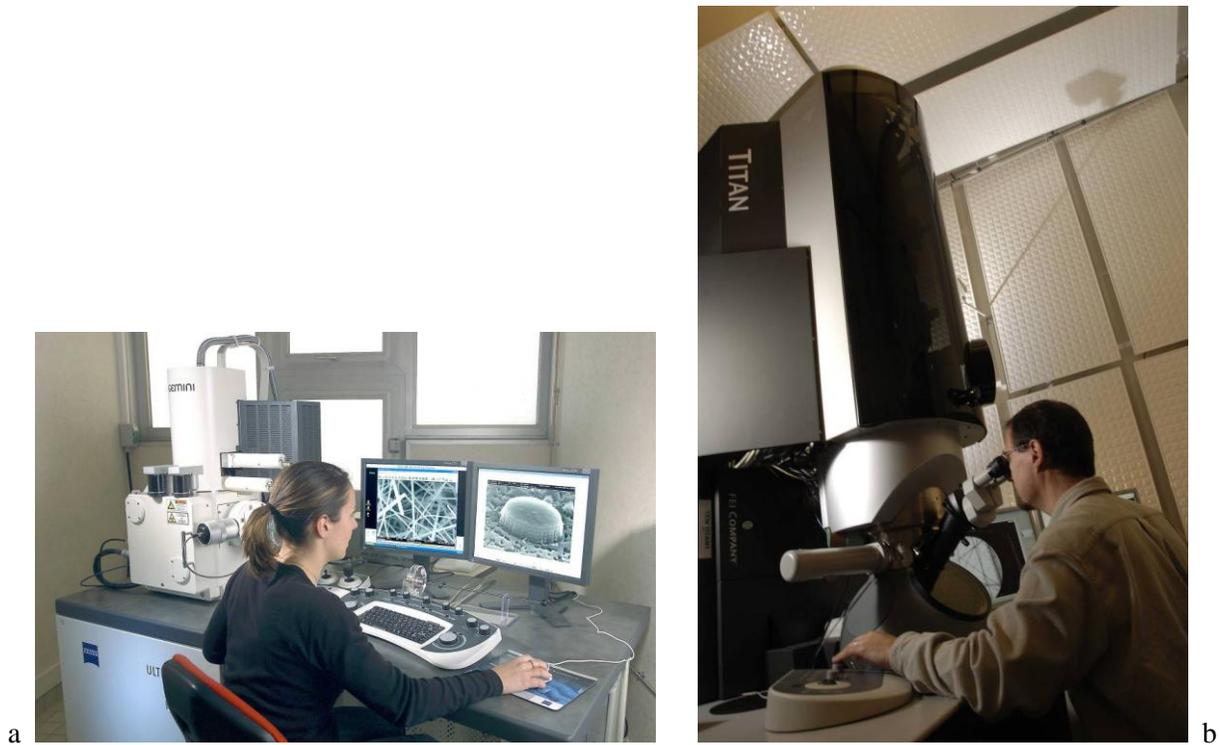

a     b

Figure 2 : A scanning electron microscope (a) and a transmission electron microscope (b) at MINATEC/ Reproduced with the kind permission of Artechnique, CEA (a) and P. Stroppa, CEA (b)

**Resonators**

Although the realization of nanowires is already an achievement, more complex micro-objects can be frabricated. As examples which has a definite application, optical micro resonators may be cited. These objects are capable of selectively retaining the photons of a certain wavelength, like macroscopic Fabry-Perot optical cavities do[1].. The electromagnetic field is then locally enhanced by a resonance effect. The quality factor $Q$[2] of the device is something like the number of times a photon crosses the cavity in both directions. Nanoresonators of Fabry-Perot can achieve a quality factor of the order of 60,000 [3]. To do better, the Fabry-Perot geometry must be replaced by another one, which corresponds to so-called 'whispering galleries'[3].. They can be for instance disks carried by nanopillars. To improve the performance of these resonators, annealing can transform the disks into tores of a remarkable perfection (Fig. 3). Quality factors are then obtained in the range of $10^7$-$10^9$ -- the highest values ever achieved in optical microstructures to date.

---

[1] The term "Fabry-Perot"corresponds to the geometry (a transparent medium between two mirrors) that is common to the micro-resonator and interferometer built by Fabry and Perot. The transparent medium is different, its thickness is different and the use is different

While these devices are of micrometric rather than nanometric size, Fig. 3 suggests that the surface is, even at the nanometer size, free from defects which would be the source of light scattering and loss of quality. The achievement of good micrometric devices implies a mastery of nanometric details and therefore belongs to nanoscience

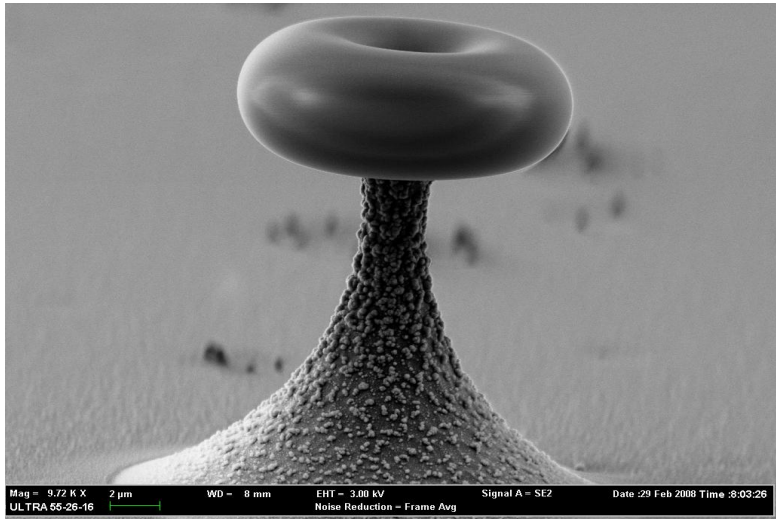

Figure 3 : A "whispering gallery" as seen by scanning electron microscopy. The torus has been "polished" by annealing. The authors are grateful to Jean Baptiste Jager (INAC/SP2M/SiNaPS) for this image.

Chemical fabrication of nanoobjets.

By chemical methods, nano-objects can be obtained in large quantities as colloids in solution. Through a sufficiently rapid injection of the appropriate reactants into a heated solvent, a rapid nucleation results, followed by particle growth. One can thus obtain nanocrystals of a few nanometers in diameter, which consist of a few hundred to several thousand atoms, also called "quantum dots" (objects in which the charge carriers are confined in all three dimensions of space). They will be formed for example by a II-VI (such as CdSe), III-V (such as InP), or I-III-VI semiconductor (such as $CuInS_2$) [4]. These nanocrystals are covered by suitable surfactant molecules (amphiphilic molecules with a polar head and apolar tail), which prevent their aggregation while allowing them to grow at high temperature. To further protect the core of CdSe, InP or CuInS$_2$ nanocrystals, it can capped by an inorganic shell, for example of zinc sulfide ZnS (Figure 4d).

---

[2] defined as the ratio $Q=\lambda/\Delta\lambda$, where $\lambda$ is the wavelength and $\Delta\lambda$ the resonance width.

[3] According to Wikipedia, "a whispering gallery is a gallery beneath a dome, vault, or enclosed in a circular or elliptical area in which whispers can be heard clearly in other parts of the building". That of Saint Paul Cathedral in London is particularly famous

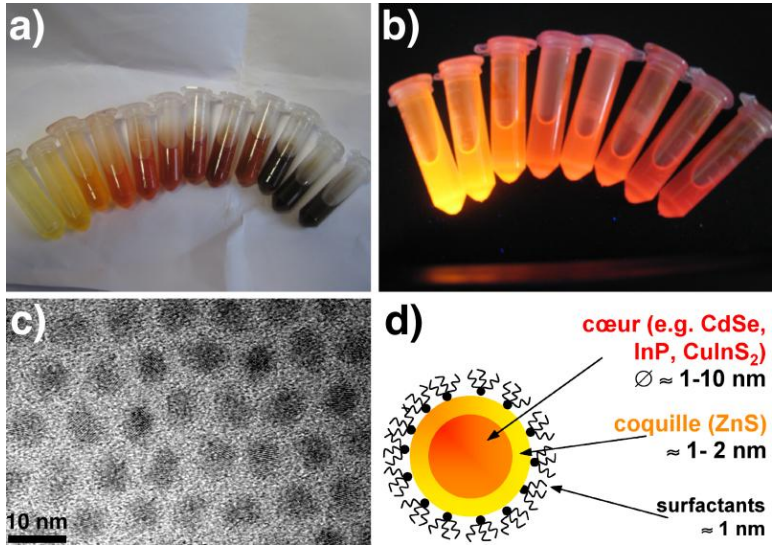

Figure 4 : CuInS2/ZnS nanocrystals of approximately 2 nm (yellow) to 4 nm (red) in diameter. a) Ambient Light b) under UV light, c) High resolution transmission electron microscopy image. d) Structure of a core / shell nanocrystal.

The diameter can be adjusted with reasonable precision by adjusting the synthesis conditions (temperature, type and concentration of reactants and surfactants,...) and suspensions can thus be obtained whose colors depend on the size of the particles (Fig. 4 a / b). Photoluminescence gives information on the average size (by the line position) and the dispersion of sizes (through the linewidth).[4]

Such quantum dots with a core/ shell structure have, when compared to organic dyes, excellent photostability, i.e. their emission and absorption properties deteriorate much more slowly in the air when exposed to light. Their fluorescence efficiency, expressed by the quantum yield, can exceed 50%.

The structure of core/ shell nanocrystals can be determined by transmission electron microscopy (Fig. 4c) which also allows analyzing the chemical composition of thin objects by electron energy loss spectroscopy (EELS).

---

[4] The emitted color corresponds to a photon whose energy is that of an exciton, i.e. close the gap of the semiconductor. This gap depends on particle size because of quantum confinement. When the size is of the same order of magnitude as the de Broglie wavelength of electrons, the energy gap increases as the nanocrystal radius decreases [5].

Researchers and applications.

The objects we describe here will probably have applications, but most of them have not yet reached the industrial stage. For instance, the resonators described above are not currently used in the instruments of our everyday life. Using them in stimulated emission, we could certainly make lasers, but lasers that read our CDs are not nanoscale. For the researchers who contribute to this article, the immediate concern is to design and manufacture new objects and study their physical properties. They think that their work will constitute the germs of new useful applications.

Nanotechnology can also be useful for fuel cells. These devices generate electricity by using combustion. The usual method to generate mechanical or electrical energy using combustion is through a heat engine, whose performance is limited by the Carnot principle. Fuel cells escape the Carnot principle, but they require a very expensive catalyst (platinum), whose life is not very long. By using catalyst particles of very small dimensions (i.e. nanoscale), we can hope to save it. On the other hand, the investigation of platinum nanoparticles by electron microscopy may lead to new ideas how to prolong the catalyst life.

Photovoltaic electricity production is another potential application of nanoparticles. The photovoltaic way is attractive for countries as Europe, which do not have many energy resources, however it is currently more expensive than other methods of energy production. Semiconductor nanocrystals, for example dispersed in suitable conjugated polymers[5] such as poly (3-hexylthiophene), are a possible solution to this situation [6]: these "hybrid" materials can be manufactured at low cost and deposited over large supports, which may even be flexible. Nanoparticles provide in that case an improvement of the light absorption by solar cells. Electron tomography, a variation of transmission electron microscopy, allows to control the structure of hybrid materials in three dimensions. The homogeneous distribution of nanocrystals in the polymer matrix is very important for the proper functioning of the solar cell.

A last example of application is the generation of white light by combining nanocrystals emitting in the blue, green and red. It might be an appreciable improvement with respect to current LEDs

Researchers and security.

The nanoparticles have a particularly strong chemical activity[6].. This makes their interest for catalysis. This can also make them toxic; for instance the silver nanoparticles, used as bactericides, are perhaps not safe for humans [7]. Another case is that of carbon nanotubes, suspected to present risks similar to those of asbestos fibers. These objects are not present in the laboratory that we visit today, but CdSe nanoparticles are produced there and they are dangerous, since both components, even in its bulk state, are toxic. Notoriously or potentially toxic nanoparticles are specifically treated in the laboratories. There is indeed a "nano-risk" as there is a risk related to chemical products handled in laboratories.

Risk is a familiar companion for experimental researchers. Chemical risks, particularly related to acids and organic solvents, are probably the most frequently present. In a large research institution, professional hazards are identified and evaluated by specialists (including physicians). At the end of the assessment procedure, a protection regulation is defined by the management, namely that of the Commissariat à l'Energie Atomique et aux Energies Alternatives (CEA) in the case of the laboratory visited to-day. Researchers have to comply with this regulation, and at the CEA the control is made by a Safety Engineer (one per department) whose exclusive task is security and by the Committee of health and safety (one per centre). This check is done here in a climate of mutual confidence which is a condition for productive research. The interest of the researcher is clearly that the risk be as low as possible. However, systematic checks are carried out. The measurements that have been made in the laboratories that we visited did not reveal any contamination by nano-particles.

---

[5] These polymers are, in the neutral (undoped) state, organic semiconductors due to their characteristic of being conjugated, i.e. their main chain consists of alternating single and double bonds

[6] Presumably related to the high proportion of atoms situated on a surface or an edge or a tip.

As a matter of fact, the labs we have presented are not among the most exposed ones. Indeed, the objects studied are present in very small quantities and usually solidary with a substrate of millimetric dimension. The nanowires are attached to the substrate on which they have grown, while the nanoparticles prepared by chemical means are not volatile (because of the organic coating) and inhalation is not to be feared. Concerning the nano-risk, appropriate means of protection are implemented, such as masks, hoods, glove boxes, specific bins and appropriate channels to collect waste.

What will happen when the nanoparticles leave the laboratory to be incorporated into a commercial product? Then it will be necessary to reassess the risks that may or may not appear, depending on the use made of these nano-objects. The relationship between the user and a product is indeed very different from the relationship between the same product and the researcher, ready to hide behind masks and hoods.

Conclusion

Nanoscience is a great conquest of the late twentieth century and early twenty-first. To make, observe and manipulate objects whose dimensions are not much larger than an atom, that is a feat which moves researchers to enthusiasm. The potential applications are very important, but the prime mover of researchers expressed here is the desire to promote knowledge. Nano-risks are taken into account taking care to neglect neither safety nor efficiency.

# Appendix. Electron microscopy

Scanning electron microscopy has been present in laboratories since 1965. It uses electrons from 1 to 40 keV. Its resolution can reach the nanometer. It is not so good as that of the scanning tunneling microscope, but its depth of field is incomparably greater (Figures 1 and 3).

Transmission electron microscopy was developed in Germany before and after the Second World War. It requires very thin samples, less than a micrometer thick. It is therefore particularly suited to the observation of nanoscale objects. With an appropriately designed transmission microscope, one can make energy-loss spectroscopy and thus perform the chemical analysis of a material. The energy loss indeed measures the energy required to extract an electron from a deep layer, which is a known feature of each atom.

Apart from energy-loss spectroscopy, specific uses of electron microscopy are electron holography and atomic scale imaging.